\documentclass[review]{elsarticle}

\usepackage{lineno,hyperref}
\modulolinenumbers[5]

\usepackage{graphicx}
\usepackage{dcolumn}
\usepackage{bm}
\usepackage{multirow}
\usepackage{textcomp}
\usepackage[]{subfigure}

\usepackage{booktabs}
\usepackage{array}
\usepackage{pgf}
\usepackage{graphicx}
\usepackage[absolute,overlay]{textpos}
\usepackage{calc}
\usepackage{amsmath}
\usepackage{amsfonts}
\usepackage{amssymb}
\usepackage{textcomp}
\usepackage{tikz}
\usepackage{gensymb}
\usepackage{color,soul}

\newcommand{\commentCWS}[1]%
{\textsf{\textcolor{blue}{#1$^{\mathrm{CWS}}$}}}
\usepackage[colorinlistoftodos]{todonotes} 

\newcommand{\figwidth}{0.9} 

\begin{document}

\begin{frontmatter}

\title{The Benefits of Trace Cu in Wrought Al-Mg Alloys}


\author[ubc]{S.~Medrano}
\author[mpie]{H.~Zhao}
\author[mpie,imperial]{B.~Gault}
\author[grenoble]{F.~De~Geuser}
\author[ubc]{C.~W.~Sinclair\corref{cor1}}
    \ead{chad.sinclair@ubc.ca}

\cortext[cor1]{Corresponding Author,}

\address[ubc]{Department of Materials Engineering, The University of British Columbia, 309-6350 Stores Road, Vancouver, Canada} 
\address[mpie]{Max-Planck-Institut f\"ur Eisenforschung, Max-Planck-Str. 1, 40237 D\"usseldorf, Germany}
\address[grenoble]{Univ. Grenoble Alpes, CNRS, Grenoble INP, SIMAP, 38000 Grenoble, France}
\address[imperial]{Department of Materials, Royal School of Mines, Imperial College London, London UK}

\begin{abstract}

The softening and strengthening contributions in pre-deformed and aged Al-Mg-Cu alloys containing 3wt.\%Mg and 0.5wt.\%Cu are evaluated by a combination of microscopy, mechanical testing and modelling.  A refined phenomenological model for the work hardening response, accounting for the separate effects of recovery and precipitation, is shown to be suitable for an unambiguous determination of the precipitation hardening contribution in these alloys. Significantly, it is found that the mechanical response of these alloys is not strongly impacted by Cu content (in the low Cu content regime), pre-deformation level or aging temperature meaning that the alloys are robust with respect to variations in composition. This is interesting from the perspective of alloy design concepts based on `recycling friendly' compositions in applications that include paint-baking.

\end{abstract}


\end{frontmatter}

\date{\today}


\section{Introduction} 

There is a growing recognition that the development of `recycling friendly' aluminum alloys will be a necessary outcome of increasing wrought alloy consumption and the corresponding increase in scrap availability \cite{raabe_strategies_2019,modaresi_component-_2014}.  The traditional approach of down-cycling wrought alloys into foundry alloys is becoming less tenable as the supply of scrap starts to outstrip the demand from casting applications \cite{modaresi_component-_2014}. One of the drivers for the increasing scrap supply is the rapid growth of aluminum in automobiles. Developing strategies that adapt existing wrought aluminum alloys (e.g. 3xxx, 5xxx and 6xxx series alloys) to tolerate, or even benefit from,  increasing levels of `impurities' (residuals) so as to achieve `recycling friendly' alloys is one strategy that appears to have promise \cite{modaresi_component-_2014}.

Wrought 5xxx series alloys are particularly challenging as `sinks' for recycled wrought aluminum alloys given that they are primarily Al-Mg alloys with only ppm of Mn and Cr added for grain size control.  For the alloy 5754, for example, Cu and Mn arriving from 3xxx and 6xxx series alloy scrap would strongly limit their use as part of the recycled scrap mix  \cite{modaresi_component-_2014}.  Finding ways for 5xxx (or 5xxx like) alloys to accommodate even small amounts of these elements would increase the ability to use a wider variety of scrap sources. 

In the particular case of Cu, there is ample evidence that small additions, small enough so as not to be detrimental to corrosion resistance\cite{engler_influence_2017}, can be beneficial in automotive 5xxx series alloys\cite{court_ageing_2002,ratchev_artificial_1999,engler_influence_2017,medrano_cluster_2018}.  Additions of less than 0.5at.\%Cu to 5XXX series aluminum alloys can reduce or even eliminate the softening experienced by stamped parts during the automotive `paint bake' cycle \cite{court_ageing_2002,ratchev_artificial_1999,engler_influence_2017}.  Given that up to 20\% of the as-stamped strength can be lost to recovery during the paint bake cycle, allowing for an increased level of Cu could create a virtuous circle; improving properties to allow designs incorporating less material, and higher scrap fractions (and lower scrap quality) in the alloy fabrication.

While it has been known for 20 years \cite{ratchev_artificial_1999} that as little as 0.2wt.\%Cu can be used to eliminate softening in 5xxx alloys during paint baking, the phenomena that control the evolution of the strength have not been clearly identified. Recent work has shown that precipitation in Al-Mg alloys containing $\sim$0.2 wt.\%Cu results in the formation of clusters/GPB zones reminiscent of those formed in conventional 2xxx series Al-Cu-Mg alloys. Importantly, however, the composition of the clusters/GPB zones in the low Cu alloys are richer in Mg and leaner in Cu compared to those in 2xxx series alloys.  This allows for a significant number density of particles and thus significant precipitation strengthening despite the relatively low Cu content \cite{medrano_cluster_2018}.  

Precipitation during the paint bake process can be expected to have two coinciding effects on the mechanical properties.  First, precipitation hardening alone could act to counterbalance the softening induced by recovery.  As noted above, a significant amount of precipitation hardening was observed in Al-Mg-Cu alloys after solution treating and aging at 160 and 200$^\circ$C \cite{medrano_cluster_2018}. In parallel, precipitation can also act to reduce the rate of dislocation loss through recovery (see e.g. \cite{roumina_recovery_2010,zurob_modeling_2002}).  Precipitation in the presence of a pre-existing dislocation network can lead to heterogeneous nucleation on dislocations, this leading to a stasis in the recovery kinetics, the duration of which is determined by the balance between the coarsening behaviour of the precipitates and the driving force for recovery (the dislocation density itself)\cite{roumina_recovery_2010}.  In the case of Al-Mg-Cu (and 2xxx series alloys as well) there is evidence that heterogeneous precipitation of metastable S-like phases on dislocations is prevalent \cite{ratchev_precipitation_1998}.  It was reported, for example, that precipitation into a pre-deformed Al-Mg-Cu alloy accelerated the formation of the S-phase at the expense of cluster/GPB zone formation \cite{ratchev_precipitation_1998}.  Similar observations have been observed for other alloys as well (see e.g. \cite{Araullo-Peters2014a,Gumbmann_2016})

\begin{figure}[htbp]
\centering
\includegraphics[width=\figwidth\textwidth]{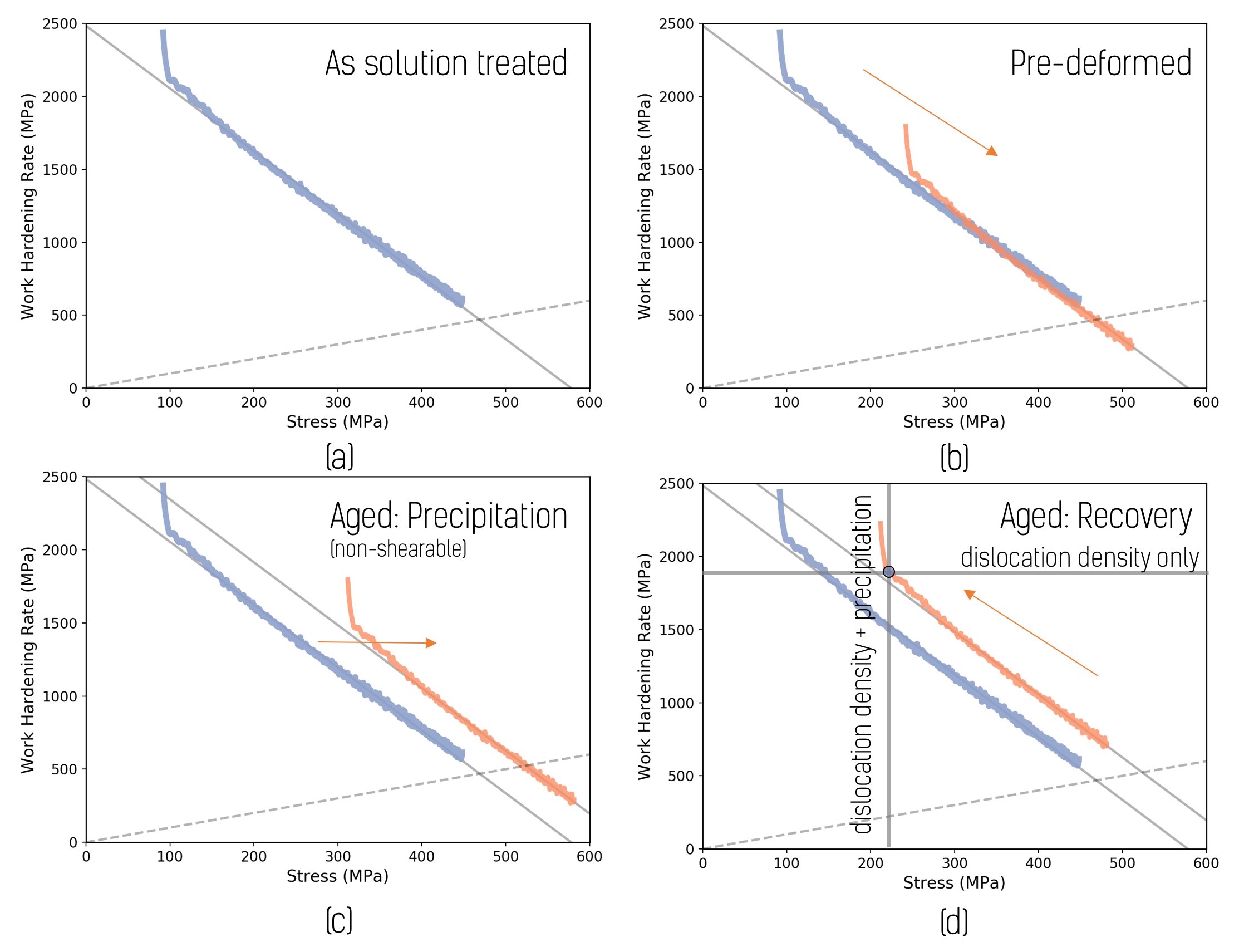}
\caption{An idealized view of the changes of flow stress and working hardening, illustrated via a Kocks-Mecking plot, for a precipitation hardenable alloy (containing shearable precipitates) during strain aging.  a) The as-solution treated (fully recrystallized) mechanical behaviour b) Upon predeformation, the dislocation density is raised causing the flow stress to increase and the work hardening rate to drop c) The addition of shearable precipitates (without recovery) would shift the flow stress without changing the work hardening rate d) recovery with precipitation would cause a lowering of the dislocation density, causing a reduction of the flow stress and an increase in the work hardening rate.  The ultimate result is that the yield stress and work hardening rate at yield could (in theory) be used to separate the effects of precipitation and recovery in a strain aged material.}
\label{fig:schemaging}
\end{figure}

Separating the effects of 1) strengthening by precipitation hardening and 2) reduction of softening by precipitation is difficult, particularly when the kinetics of recovery and precipitation coincide.  One possible way to approach this problem, using mechanical property data alone, would be to recognize that the two phenomena can have distinctly different effects on different parts of the stress-strain response (see Figure \ref{fig:schemaging}). In the case of precipitation hardening by small, shearable precipitates the effect is fully attributable to the yield strength, the change in the work hardening rate being negligible.  In contrast, softening by recovery (through the loss of dislocations) would have an effect on both the yield strength \emph{and} work hardening rate.  This idealized view is illustrated in Figure \ref{fig:schemaging}.  If we take a solution treated alloy and pre-deform it, increasing its dislocation density, we find that the pre-deformed material when subsequently tested will exhibit a work-hardening behaviour that falls along the same line as the original solution treated alloy (Figure \ref{fig:schemaging} b).  Precipitation hardening by shearable precipitates will, in contrast, shift the curve for the material to the right without changing its slope (Figure \ref{fig:schemaging} c).  Recovery, will reduce the dislocation density, causing the behaviour to shift back along the work hardening curve\footnote{This is true, to a good approximation, when the properties are measured without strain-path change between pre-deformation and post aging measurements. Introducing a strain path change introduces further complexity not considered here.}(Figure \ref{fig:schemaging} d).   Ultimately, for a pre-deformed sample that has been aged under conditions where both recovery and precipitation have taken place, prediction of the work hardening rate (dependent on dislocation density only) and flow stress at yield (dependent on dislocation density and precipitation hardening) can be used to identify both contributions (Figure \ref{fig:schemaging} d).  

In this work, we start from the concept outlined above with the aim of identifying the individual contributions of recovery and precipitation to the mechanical properties of a pre-deformed and aged Al-3\%Mg alloy containing 0.5wt.\%Cu (i.e. Al-3.33at.\%Mg-0.21at.\%Cu). This builds on our recent work using the same alloy, where we have characterized the precipitation sequence, kinetics and hardening in the absence of pre-deformation\cite{medrano_cluster_2018}.  Here, we use low temperature tensile tests to introduce pre-deformation as well as to characterize the flow stress and work hardening response following aging at 160$^\circ$C and 200$^\circ$C. Illustrative atom probe tomography (APT) has been employed to circumvent the challenge of characterizing precipitates with low atomic number contrast (e.g. those containing neighbouring elements like Al and Mg) while allowing for characterization of segregation on defects. We will show that the idealized analysis shown in Figure \ref{fig:schemaging} is too simple to be fully useful but that the desired separation of flow stress contributions can ultimately be recovered using a more sophisticated analysis of the work hardening behaviour. Based on this analysis the relationship between pre-deformation level, Cu alloy content and strength is proposed.  This work highlights the possibility of adjusting the thermomechanical processing strategy based on minority (residual) element levels to give desirable properties using recycling friendly alloy compositions.

\section{Materials and Experimental Method}
A laboratory cast and hot rolled 5 mm sheet was provided for this study by the Novelis Global Technology Centre (see Table \ref{tab:comp}).  This material was cold rolled to a thickness of 1 mm (80\% reduction in thickness). Tensile samples were cut and solution treated at 550$^\circ$C for 10 min, then quenched immediately in water.  
These samples were then deformed in tension to introduce deformation prior to aging. These tests were performed at 77 K, at a nominal strain rate of 10$^{-3}$ s$^{-1}$. These conditions were selected so as to avoid the complicating effects of dynamic strain aging \cite{fazeli_modeling_2008} as well as to reduce solute mobility during deformation. Deformation was performed to raise the flow stress by $\Delta\sigma =$ 44 MPa or 176 MPa, corresponding to $\sim$ 2 and 10\% tensile strain respectively.  The samples were then stored in liquid nitrogen until ready to be aged. Aging of the pre-deformed samples was conducted at 160$^\circ$C or 200$^\circ$C in a stirred oil bath.  Following aging, samples were immediately tested in tension following the same procedure as used for pre-deformation.

\begin{table*}[htbp]
\resizebox{\textwidth}{!}{%
 \centering
 \begin{tabular}{ccccccccccc}
  \toprule
   & & \textbf{Mg} & \textbf{Cu} & \textbf{Fe} & \textbf{Si} & \textbf{Ti} & \textbf{Ni} & \textbf{Mn} & \textbf{Zn} & \textbf{Cr}  \\
  \midrule
			& (\small{wt.\%}) & 2.90 & 0.54 & 0.1 & 0.05 & 0.015 & 0.007 & 0.001 & 0.002 & 0.001 \\
			
  \bottomrule  
 \end{tabular}}
 \caption{Composition of studied Al-3Mg-0.5Cu alloy obtained by optical emission spectroscopy.  Balance is aluminum.}
 \label{tab:comp}
\end{table*}

Specimens for transmission electron imaging in a scanning electron microscope (STEM-in-SEM) and APT were cut from the gauge section of pre-deformed and aged samples. STEM samples were prepared by grinding the sample to reduce its thickness, followed by punching out 3~mm diameter discs.  These disks were electropolished in a nitric acid methanol solution to perforation. Samples were observed using a FEI Versa dual beam microscope at 30 keV, with the STEM detector operating in bright field mode.  

For APT, needle-shaped samples were prepared via a 2-stage electrochemical polishing technique \cite{miller_atom_2000,medrano_cluster_2018}.  Data was collected using a Cameca LEAP 5000 XS.  Experiments were performed using a base temperature of 50K, in high-voltage pulsing mode, with 20\% pulse fraction. The DC voltage was adjusted to maintain a detection rate of 1 ion per 100 pulses. Data reconstruction was performed in the commercial software package IVAS 3.6.14 using the calibration methods outlined elsewhere \cite{gault_estimation_2008,gault_advances_2009}. Isocomposition surfaces have been used to highlight regions of enhanced solute (segregation, precipitation) \cite{hellman_efficient_2003} and the composition of individual objects was approximated by composition profiles across features of interest. 

\section{Experimental Characterization of Strain Aging Response}

\subsection{Yield Strength and Microstructure Evolution}
\begin{figure}[htbp]
\centering
\includegraphics[width=\figwidth\textwidth]{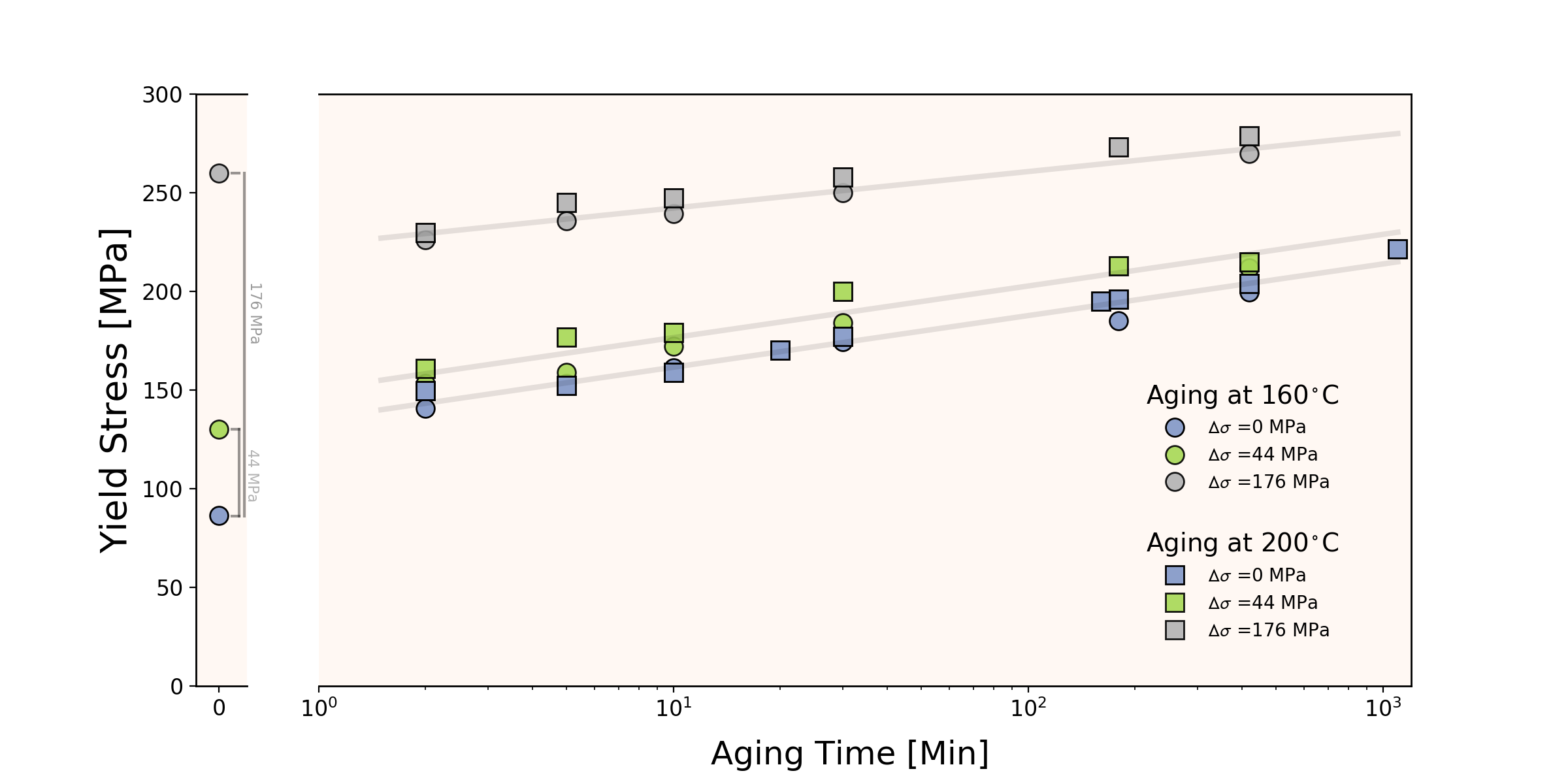}
\caption{The evolution of the 0.2\% offset yield strength aging at 160 and 200$^\circ$C and for the as-solution treated sample as well as samples pre-deformed to $\Delta\sigma = 44$ MPa and $\Delta\sigma = 176$ MPa.  The yield strength of the solution treated sample and as-deformed samples is shown on the far left side of the plot.  Lines are drawn as a guide to the eye.}
\label{fig:yieldstrength}
\end{figure}

As a first insight into the aging response, Figure \ref{fig:yieldstrength} shows the evolution of the yield strength for three conditions; the material with no pre-deformation and the two pre-deformed cases aged at 160$^\circ$C and 200$^\circ$C. In the case of aging starting from the as-solution treated material, it has been shown \cite{medrano_cluster_2018} that the hardening is the result of increasing cluster/GPB zone size.  For the material pre-deformed to $\Delta\sigma = 44$ MPa ($\sim$ 2\% strain) the evolution of the yield stress is nearly the same as that found when aging starts from the solution treated material, the offset between the two sets of data being approximately 15 MPa.  In the case of the material pre-deformed by $\Delta\sigma = 176$ MPa ($\sim$ 10\% strain) an initial drop in yield stress is observed if one compares the as-deformed yield stress to the yield stress after 2 minutes of aging.  Further aging, however, leads to continuous strengthening, the rate being very close to that of the material starting from the solution treated state and the samples pre-deformed to $\Delta\sigma = 44$.  The aging temperature (160$^\circ$C or 200$^\circ$C) makes no significant change in the response for the pre-deformed samples, this being consistent with what was shown for a range of alloy compositions for the as annealed material \cite{medrano_cluster_2018}. 

The similarity of the increase in yield strength for all of the conditions at times longer than 2 min suggests that recovery dominates the flow stress at short times (particularly for $\Delta\sigma = 176$ MPa) but that precipitation hardening dominates the yield strength evolution at longer times.  In the case of the solution treated and aged samples, it has been shown that the hardening is dominated by a high density of small clusters/GPB zones, similar in morphology and density to those found in 2xxx series alloys but containing much lower Cu contents \cite{medrano_cluster_2018}.  It was shown that the volume fraction of these features did not change significantly over the aging times shown in Figure \ref{fig:yieldstrength} but that the size of the particles did, this leading to the observed strengthening.  

\begin{figure}[htbp]
\centering
\includegraphics[width=\figwidth\textwidth]{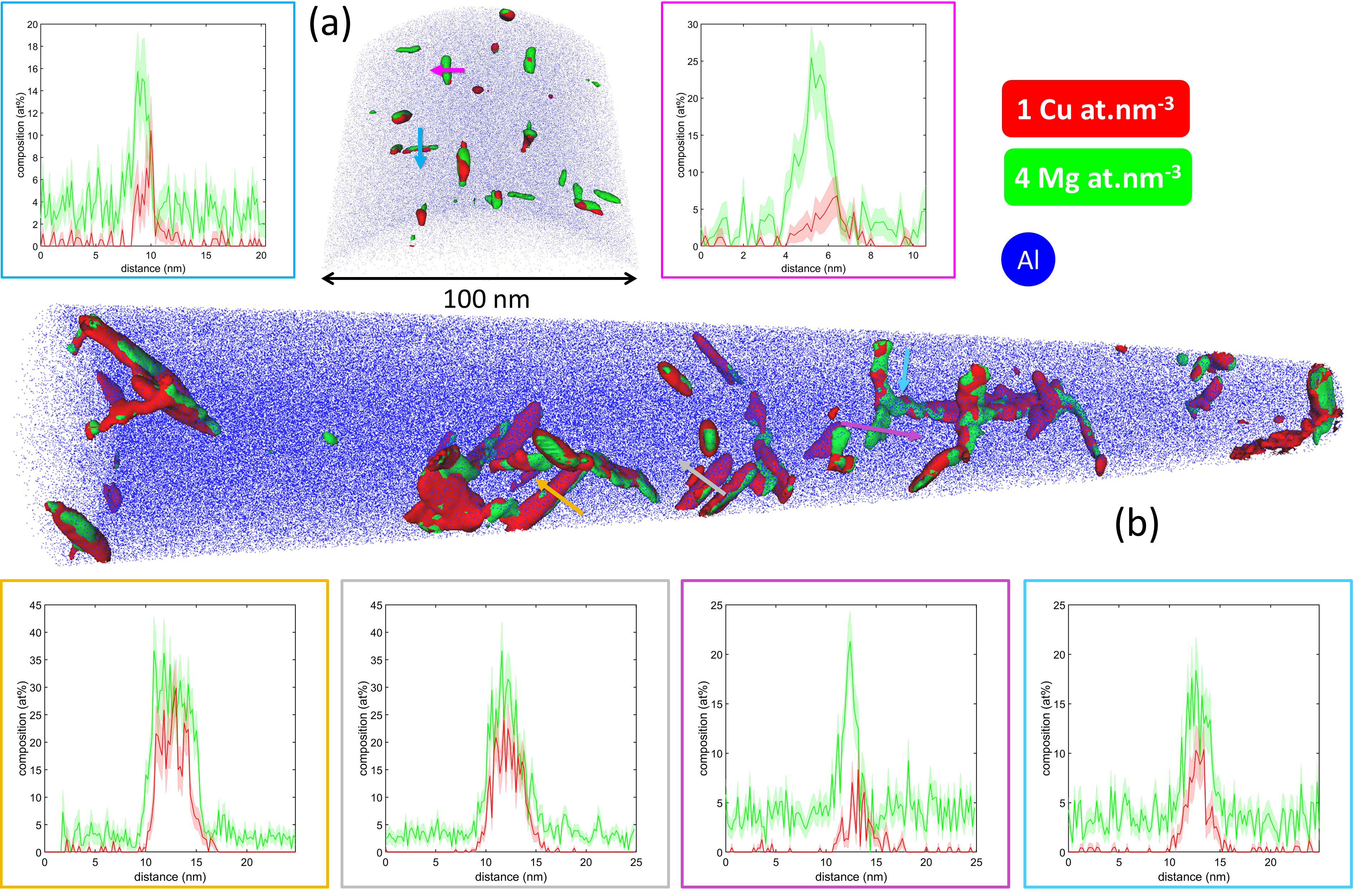}
\caption{APT reconstruction of specimens prepared from samples solution treated then a) aged for 160 min at 200$^\circ$C (no pre-deformation)  and b) pre-deformed to 44 MPa then aged for 160 min at 200$^\circ$C.  Iso-concentration surfaces are superimposed on the data to highlight the presence of precipitates. Both reconstructions are displayed with the same scale. One-dimensional composition profiles for some of the particles found in each of the data sets are displayed alongside the reconstruction, and for each, an arrow of the same colour is included in the reconstruction to indicate which microstructural object was interrogated. The shaded area corresponds to the $2\sigma$ of the counting statistics in each of the 0.2 nm bin of the profile.}
\label{fig:apt}
\end{figure}

Figure \ref{fig:apt} shows illustrative atom probe tomography analyses for two samples aged at 200$^\circ$C for 160 min with a) no pre-deformation and b) pre-deformation to 176 MPa \footnote{A more detailed analysis of samples aged with no-predeformation can be found in \cite{medrano_cluster_2018}}. The tomographic reconstructions are displayed at the same scale. For both, sets of iso-contour plot are superimposed onto the point cloud (only approx. 1\% of the detected Al ions are shown), in order to highlight the population of clusters/GPB zones. 
In the aged sample 3 variants of small, elongated precipitates are visible. One-dimensional composition profiles calculated in 5-nm-diameter cylinders positioned perpendicular to the long axis of the feature (indicated by coloured arrows) are also shown. The particles are mostly Mg-rich, reaching typically close to 25 at.\% Mg, with 2-5 at \% Cu (i.e. 22 wt.\% Mg and 5-6.5 wt.\% Cu).  

In contrast, for the pre-deformed material, the distribution of the solute elements is much more heterogeneous, both spatially and compositionally, as highlighted by the iso-contours showing solute rich regions. The geometry of the features suggests segregation and/or precipitation along dislocations. Indeed, some regions along these features appeared to have Mg:Cu ratios close to those observed for clusters/GPB zones in the solution treated and aged samples (e.g. Figure \ref{fig:apt}a) while others had a Mg:Cu ratio closer to 1. Generally, the particles are larger and less well dispersed in the pre-deformed and aged sample compared to the sample aged directly from the as-solutionized state.  

\begin{figure}[htbp]
\centering
\includegraphics[width=\figwidth\textwidth]{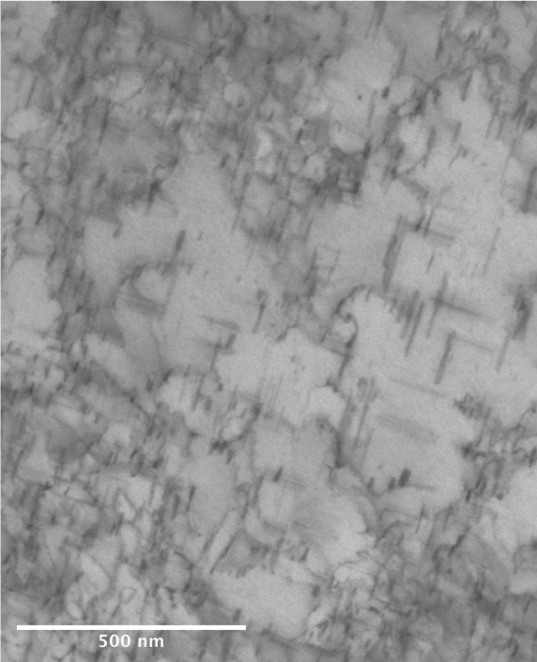}
\caption{Bright field STEM-in-SEM image from a sample pre-deformed to $\Delta\sigma$=44 MPa and aged at 200$^\circ$C for 1600 min illustrating the formation of S-phase laths preferentially along dislocations.}
\label{fig:STEM}
\end{figure}

Aging of pre-deformed samples leads to the formation of S-phase precipitates preferentially along dislocations, by-passing the more typical sequence of clusters/GPB zone formation followed by S-phase formation \cite{ratchev_precipitation_1998}. A similar process has also been reported for other alloys, including an Al-Cu-Li alloy \cite{Araullo-Peters2014a}. This is visible in samples prepared here, where lath-like S-phase can be seen along dislocations in Figure \ref{fig:STEM}. The composition of some of the larger particles imaged by APT in Figure \ref{fig:apt}b is consistent with the S-phase as well. 

The microstructural differences between the solution treated+aged and pre-deformed+aged samples belies the very similar rates of hardening shown in Figure \ref{fig:yieldstrength}.  While in our previous work on the hardening response of solution treated and aged samples we were able to use information directly from APT experiments to parameterize a precipitation hardening model, here the complexity of the segregation/precipitation in the pre-deformed samples does not allow for such an approach.  Rather, to more quantitatively evaluate the precipitation hardening  we can look at the work hardening behaviour (Figure \ref{fig:stressstrain_workhardening}).  

\subsection{Work Hardening Analysis: Separating the Effects of Recovery Induced Softening  and Precipitation Hardening}
\begin{figure}[htbp]
\centering
\includegraphics[width=\figwidth\textwidth]{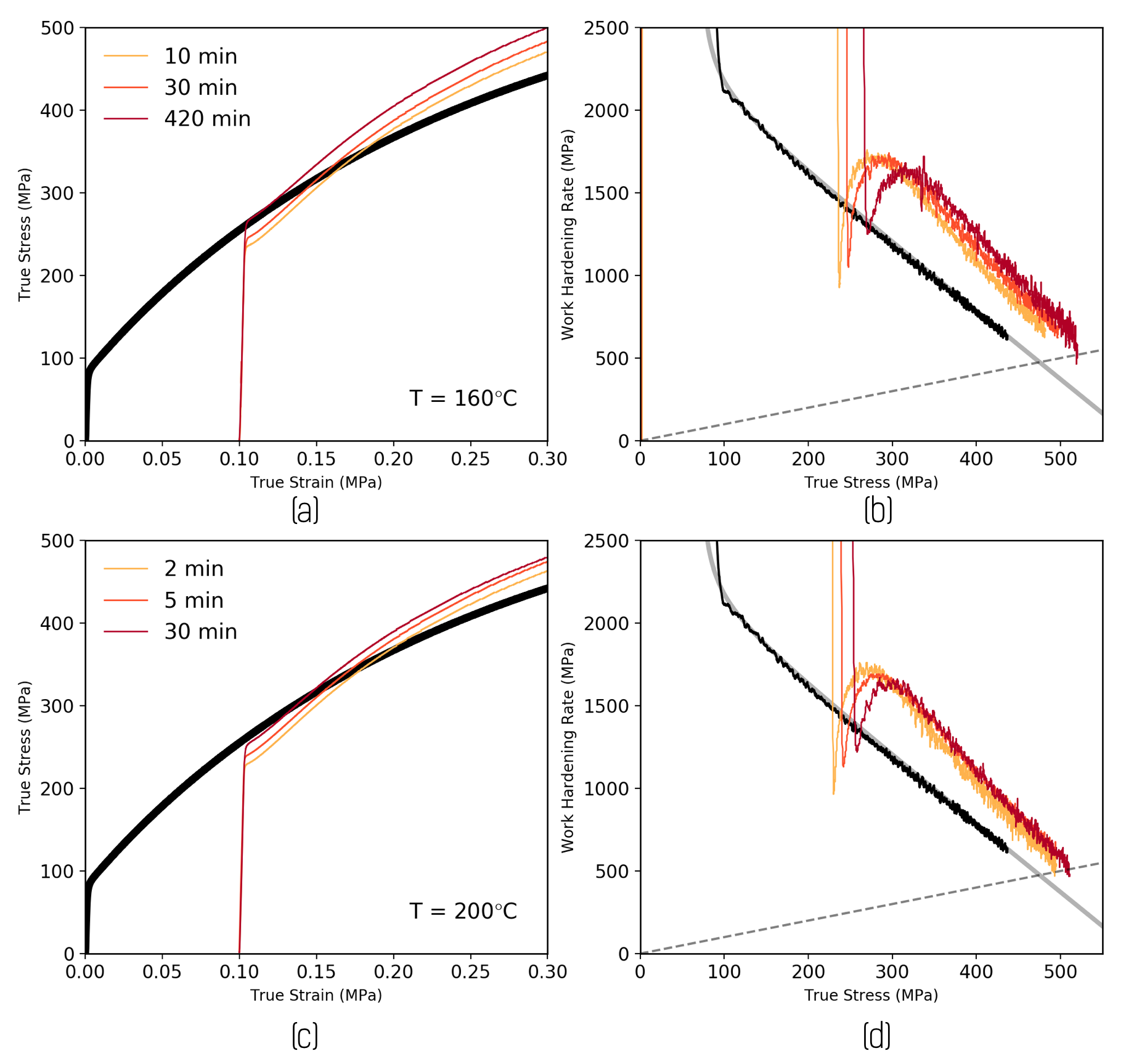}
\caption{Evolution of the stress-strain (a,c) and work-hardening behaviour (b,d) of a sample pre-deformed to $\Delta \sigma$ = 176 MPa.}
\label{fig:stressstrain_workhardening}
\end{figure}

A cursory observation of Figure~\ref{fig:stressstrain_workhardening} shows that the real response of the alloy is more complex than the idealized response illustrated in Figure~\ref{fig:schemaging}.  The most noticeable discrepancy between Figures~\ref{fig:schemaging} and \ref{fig:stressstrain_workhardening} is the (transient) low work hardening rate ($d\sigma/d\epsilon$) at the onset of yielding. A more careful inspection of Figure~\ref{fig:stressstrain_workhardening} also shows that, unlike the idealized response, the work hardening vs. flow stress curves for the as-solution treated and pre-deformed and aged samples are not parallel to one another.  Rather, it appears that parallelism is approached asymptotically at high flow stresses. This is particularly notable in Figure~\ref{fig:stressstrain_workhardening}d.

Very similar behaviour has been recently reported and discussed for a non-precipitation hardenable Al-Mg alloy pre-deformed and aged under very similar conditions to those used here \cite{medrano_transient_2020}. Therein, a simplified version of the Kubin and Estrin \cite{kubin_evolution_1990} two-internal state variable model was developed. In that case, it was argued that solute-dislocation interactions amplified by the low temperature aging treatment led to changes in the static and dynamic recovery response of forest dislocations.  Here, we adapt this model to the case of Al-Mg-Cu alloys. The precipitates are considered as shearable particles and we retain the basic concept outlined in Figure~\ref{fig:schemaging} as our method for separating the effects of recovery and precipitation hardening on the yield and flow stress. Here we present the most salient aspects of the model, including the modifications to include the effect of precipitation hardening.  We start by writing the flow stress as a function of the strain rate and forest dislocation density ($\rho_f$).

\begin{equation}
    \sigma = \sigma_{0} + \sqrt{\sigma_{ss}^2+\sigma_{ppt}^2} + M\alpha G b\sqrt{\rho_f+\rho_f^*}\left(\frac{\dot{\epsilon}}{\dot{\epsilon}_0}\right)^m
    \label{eqn:taylor2}
\end{equation}

In this expression $\sigma_0$ is a friction stress including effects of lattice resistance and the grain size. The next two terms $\sqrt{\sigma_{ss}^2+\sigma_{ppt}^2}$ are the contributions of solid solution strengthening and precipitation strengthening, these being added based on the density of obstacles based on the expectation that the contributions from these two terms will be similar \cite{medrano_cluster_2018}. 

The reference strain rate, $\dot{\epsilon}_0$ is assumed to be proportional to the mobile dislocation density $\dot{\epsilon}_0 = \beta \rho_m$ and was assumed to evolve as,

\begin{equation}
    \frac{1}{\rho_m^s}\frac{d\rho_m}{d\epsilon} = \eta\left(1-\frac{\rho_m}{\rho_m^s}\right)
    \label{eqn:rhom}
\end{equation}

Where $\eta$ is a parameter related to the rate of mobile dislocation density generation and $\rho_m^s$ is a saturation mobile dislocation density. The mobile dislocation density will rise with strain to $\rho_m^s$ corresponding to $\dot{\epsilon_0} = \dot{\epsilon}$, or $\rho_m^s = \beta^{-1}\dot{\epsilon}$. Note that in eqn.~\ref{eqn:rhom} one only needs to identify the ratio $\rho_m/\rho_m^s$, not the absolute value of $\rho_m$.   

In equation \ref{eqn:taylor2}, two separate forest dislocation densities are introduced, $\rho_f$ and $\rho_f^*$.  During pre-deformation at low temperature we expect to generate forest dislocations $\rho_f$, which have interacted with approximately immobile solute (due to the low temperature of pre-deformation).  Upon aging, two processes will be activated.  First, some of these forest dislocations will be lost due to static recovery.  At longer times, we anticipate some fraction of these dislocations to be modified due to solute segregation to them and/or precipitation on them. These phenomena are anticipated to change both the rate of static recovery as well as the rate of dynamic dislocation storage and annihilation on reloading following aging. The density of dislocations identified by $\rho_f^*$ are those that have been decorated by solute (or precipitates) during artificial aging.  The total forest dislocation density and the end of aging ($\rho_f+\rho_f^*$) is expected to be less than that at the end of pre-straining due to recovery.  

The low temperature deformation after aging is expected to only reduce $\rho_f^*$, this having been taken in \cite{medrano_transient_2020} to follow,

\begin{equation}
    \frac{d\rho_f^*}{d\epsilon} = - k_4^*\rho_f\left[1-\exp{\left(-\sqrt{\frac{\rho_f^*}{\rho_f}}\right)}\right]
    \label{eqn:rateofloss}
\end{equation}

The important concept here is that the density $\rho_f^*$ recovers (dynamically) more slowly than those dislocations generated by deformation ($\rho_f$), this leading to an enhanced rate of work hardening over the initial portion of the stress-strain curve.

During deformation, the evolution of $\rho_f$ is assumed to obey,

\begin{equation}
 \frac{d\rho_f}{d\epsilon} = k_2\frac{\rho_m}{\rho_m^s} + k_3\sqrt{\rho_f} +k_3^*\sqrt{\rho_f^*} - k_4\rho_f
    \label{eqn:rhof2}
\end{equation}

where different rates of storage are attributed due to the interaction with `solute loaded' and `solute unloaded' forest dislocations. 

In the above system of equations, there are several parameters that need to be set to predict the mechanical properties.  In the case of a solution treated sample it is assumed that $\rho_f^*$ can be approximated as zero meaning that eqn.~\ref{eqn:rhof2} depends only on $k_2$, $k_3$ and $k_4$.  In the limit where $\rho_m = \rho_m^*$ then this further simplifies as the first term in eqn.~\ref{eqn:rhof2} becomes a constant, this reducing to the classic Kocks-Mecking analysis (Figure~\ref{fig:schemaging} with an initial non-linearity (at low stress) controlled by the value of $k_2$ . 

The value of $\sigma_{ss}$ can be obtained directly from the alloy content of the sample, the value used here being the same as the value used in \cite{medrano_cluster_2018}.  One complicating factor is the fact that precipitation can act to deplete solute from the matrix, this making $\sigma_{ss}$ a function of the aging condition.  In our work on aging starting from the as-solution treated state \cite{medrano_cluster_2018} it was shown that the solid solution strengthening was dominated by Mg and that the amount of Mg and Cu found in clusters/GPB zones was insufficient to lead to a significant change in $\sigma_{ss}$.  Here, for simplicity, we also assume that $\sigma_{ss}$ can be taken to be constant, independent of the aging condition.   

Taking the above points into consideration, a single test on a solution treated sample, allows one to establish values for $\sigma_0$, $k_2$, $k_3$ and $k_4$, the values found for the alloy studied here being given in Table \ref{tab:params}.

\begin{table}[htbp]
 \centering
 \begin{tabular}{ccc}
  \toprule
  \textbf{Parameter} & \textbf{Value} & \textbf{Comment}  \\
  \midrule
  \multicolumn{3}{c}{Material Properties} \\
   \midrule
  $b$ & 0.4 nm & Burgers vector  \\
  $\mu$ & 25 GPa & Shear modulus  \\
  $M$ & 3 &  Taylor factor \\
  $\alpha$ & 0.3 & Saada constant  \\
  $m$ & 0.01 & Rate sensitivity  \\
  \midrule
  \multicolumn{3}{c}{Hardening Law Parameters (Solutionized)}   \\
  \midrule
  $\sigma_0$ & 5 MPa & \small{Friction stress and grain size} \\
 $\sigma_{ss}$ & 65 MPa & \small{Solid solution strengthening from \cite{medrano_cluster_2018}}  \\
 $\rho_f^0$ & $2\times10^{12}$ m$^{-2}$ & \small{From $\sigma_0$ and experimental yield stress}  \\
  $k_2$ & $22\times10^{13}$ m$^{-2}$ & \small{Non-linear response at low stress} \\
  $k_3$ & $5.7\times10^{8}$ m$^{-1}$ & \small{Asymptotic linear fit at high stress} \\
  $k_4$ & 8.4 & \small{Asymptotic linear fit at high stress} \\
  \midrule
  \multicolumn{3}{c}{Hardening Law Parameters (Pre-deformed $+$ Aged)}\\
  \midrule
  $\eta$ & 50 & generation rate of $\rho_m$ \\
  $k_{3}^*$ & $2.3\times10^{8} m^{-1}$ & Taken to be equal to $0.4k_3$ \\
  $k_{4}^*$ & 4.2 & Taken to be equal to $0.5k_4$\\
  \bottomrule  
 \end{tabular}
 \caption{Parameters used in modelling the work hardening response using equations \ref{eqn:taylor2}, \ref{eqn:rhom}, \ref{eqn:rhof2}. }
 \label{tab:params}
\end{table}

The remaining parameters in eqn.~\ref{eqn:rhof2}, $k_3^*$ and $k_4^*$, were obtained as a single best fit to all of the pre-deformed and aged conditions used in this study. Following the findings in \cite{medrano_transient_2020}, we restrict the values of $k_3^*$ and $k_4^*$ to be of the same order of magnitude as $k_3$ and $k_4$.   The remaining global parameter to identify is $\eta$ in eqn.~\ref{eqn:rhom}. As was shown in \cite{medrano_transient_2020} the exact value is not sensitive to the resulting fit, therefore this parameter was taken to be identical to the value used in \cite{medrano_transient_2020}. 

The remaining parameters that are required, and which depend on the level of pre-strain and the aging time/temperature are, i) the density of forest dislocations following aging ii) the ratio of the density of mobile dislocations ($\rho_m/\rho_m^s$) following aging and iii) the precipitate contribution to the flow stress, $\sigma_{ppt}$.  The density of dislocations (mobile and forest) following pre-straining are estimated from the fit to the solution treated stress-strain curve at $\Delta\sigma = 176$~MPa. Following aging, the density of mobile dislocations is expected to be reduced as some mobile dislocations are locked and converted to forest dislocations while others are lost due to recovery.  In the case of the forest dislocation density following aging, we need to distinguish between solute locked forest dislocations ($\rho_f^*$) and non-solute locked forest dislocations ($\rho_f$).  Owing to recovery, we expect that $\rho_f$ + $\rho_f^*$ will be less than the density of forest dislocations present at the end of pre-straining.  

As described in \cite{medrano_transient_2020}, the value of $\rho_m$ following pre-straining and aging affects mainly the value of the (low) work hardening rate at the onset of yielding and does not significantly impact the stress-strain response beyond $\sim$ 1\% strain.  Indeed, according to eqn.~\ref{eqn:rhom}, mobile dislocations are rapidly generated with the onset of plasticity for the parameters used here.  The value of $\sigma_{ppt}$ only impacts on the yield strength after aging while the values of $\rho_f$ and $\rho_f^*$, have a significant impact on both the yield strength and work hardening rate.  

The values of $\sigma_{ppt}$ and $\rho_f + \rho_f^*$ can be established by fitting to the yield strength of the material as well as the initial (high) work hardening rate following the initial (transient) minimum in the work hardening rate. A more direct estimate of $\sigma_{ppt}$ can be obtained from the flow stress-work hardening relationship at high flow stress.  At high flow stress, the effect of $\rho_f$ dominates that from $\rho_f^*$ leading to the recovery of a linear relationship between flow stress and work hardening rate, as obtained in a classical Kocks-Mecking analysis.  As pointed out before, this behaviour is clear in Figure~\ref{fig:stressstrain_workhardening}d where the solution treated and aged samples are seen to become parallel at high stress. 

Under these conditions, $\sigma_{ppt}$ is recovered unambiguously from the separation between the parallel solution treated and aged work hardening curves, as illustrated in the case of the simplified analysis presented in Figure~\ref{fig:schemaging}.  Knowing $\sigma_{ppt}$, one can then use the yield strength to obtain $\rho_f + \rho_f^*$.  To separate these two values, one can then use their distinctly different effect on the work hardening response.  

To capture the results found here, it was found necessary to take $\rho_f^* >> \rho_f$ following aging.  This would suggest that nearly all dislocations that do not recover during the aging treatment are segregated to and/or precipitated on, a result that would seem intuitive given the similarity of the segregation profile to split edge and screw dislocations in Al-Mg alloys \cite{dontsova_solute_2015}.  

\begin{figure}[htbp]
\centering
\includegraphics[width=\figwidth\textwidth]{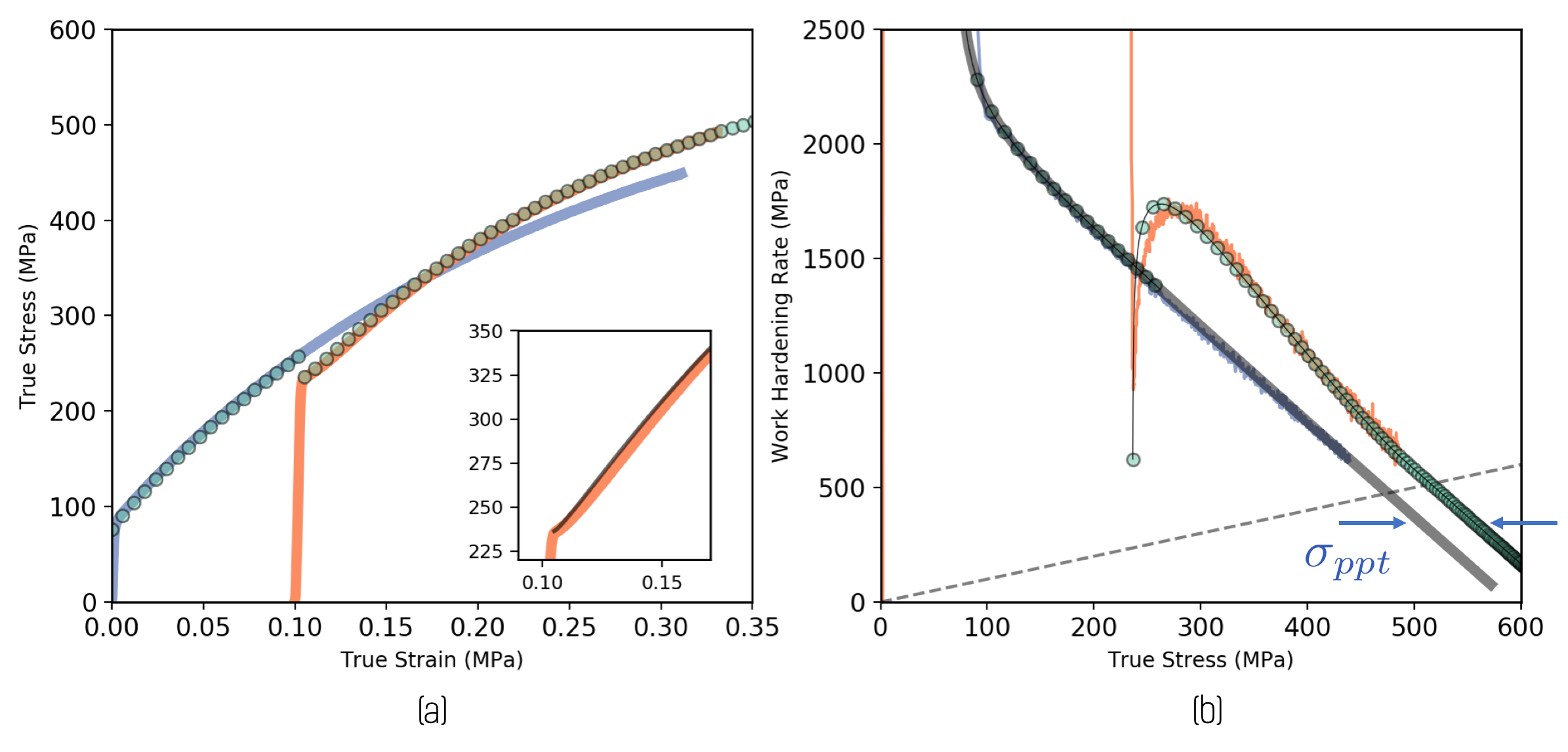}
\caption{Illustration of the fit of the model (eqns. \ref{eqn:taylor2}, \ref{eqn:rhof2} and \ref{eqn:rhom}) to a) the stress-strain response of the solution treated material (blue) and the material pre-strained then aged for 10 minutes (orange).  Experimental data are shown as lines while the model fits are shown as circles. Shown in (b) is the flow stress-work hardening relation for the same materials.  Here it is shown how the experimental data and model prediction lead asymptotically to linear behaviour at high flow stresses and how this can be used to deduce the magnitude of $\sigma_{ppt}$.}
\label{fig:illustrate_fitting}
\end{figure}

Figure \ref{fig:illustrate_fitting} shows experimental data and modelling fits for the material that had been pre-deformed by $\Delta\sigma = 176$ MPa and aged for 10 minutes at 200$^\circ$C.  Here one can explicitly see how, at high flow stresses, the work hardening rates of the solution treated and aged materials become parallel, the separation between the two giving the value of $\sigma_{ppt}$.

\begin{figure}[htbp]
\centering
\includegraphics[width=\figwidth\textwidth]{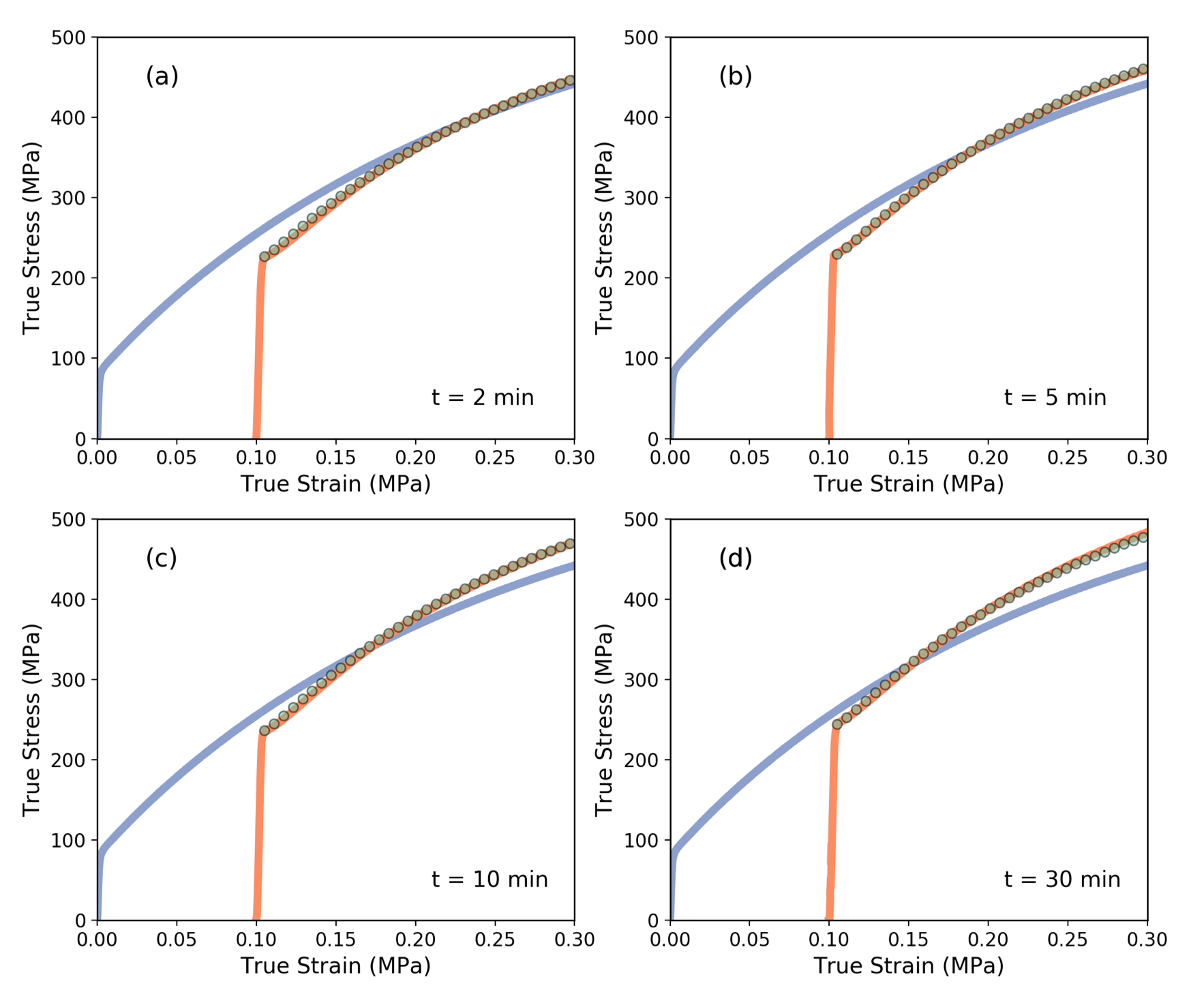}
\caption{Illustration of the fit of the model to the stress-strain response following aging at 200$^\circ$C for the indicated times.  Experiments are solid lines (blue = solution treated, orange = aged) and symbols are model predictions.}
\label{fig:fitto200C}
\end{figure}

Following the approach outlined above, we have deduced a best fit to the stress-strain response of the materials pre-deformed to $\Delta \sigma$ = 176 MPa and aged at 160$^\circ$C and 200$^\circ$C for various times.  Figure \ref{fig:fitto200C} shows the experimental stress-strain curves (solid lines) along with the fits of the model (symbols).  

\begin{figure}[htbp]
\centering
\includegraphics[width=\figwidth\textwidth]{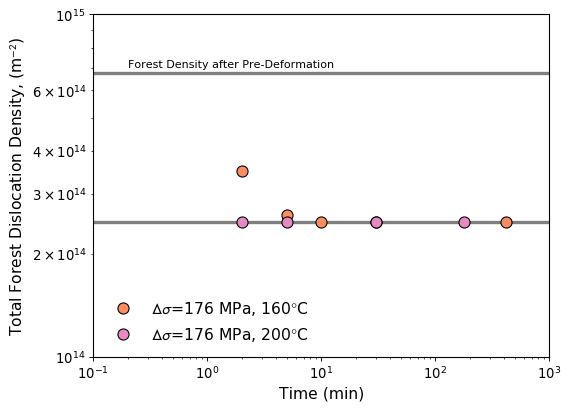}
\caption{The estimated total forest dislocation ($\rho_f + \rho_f^*$) following pre-deformation and aging. }
\label{fig:recovery}
\end{figure}

As can be seen, the fits reproduce the stress-strain response very well, the two main parameters that need to be adjusted with aging time being the forest dislocation densities directly after aging ($\rho_f$ and $\rho_f^*$) and the precipitation contribution to the strength, $\sigma_{ppt}$.  It was found that the combined best fit values of $\rho_f+\rho_f^*$ were nearly identical for all but the two shortest annealing times at 160$^\circ$C (Figure \ref{fig:recovery}).  This would suggest that recovery happens much more quickly than precipitation, reducing the forest dislocation density to $\sim$ 37\% the value that was present directly after reloading.  

The fact that further recovery seems not to take place may be a consequence of solute segregation and precipitation on those dislocations that had not recovered earlier.  This should be contrasted with the results obtained from applying the same model to very similar experimental results for a binary (non-precipitation hardening) Al-Mg alloy \cite{medrano_transient_2020} where it was observed that recovery continued obeying classic Bailey-Orowan recovery kinetics.  These results would be in line with the view presented in the case of Al-Mg-Sc alloys where a stasis in the recovery was observed owing to dislocation/precipitate interactions \cite{roumina_recovery_2010}. 

\begin{figure}[htbp]
\centering
\includegraphics[width=\figwidth\textwidth]{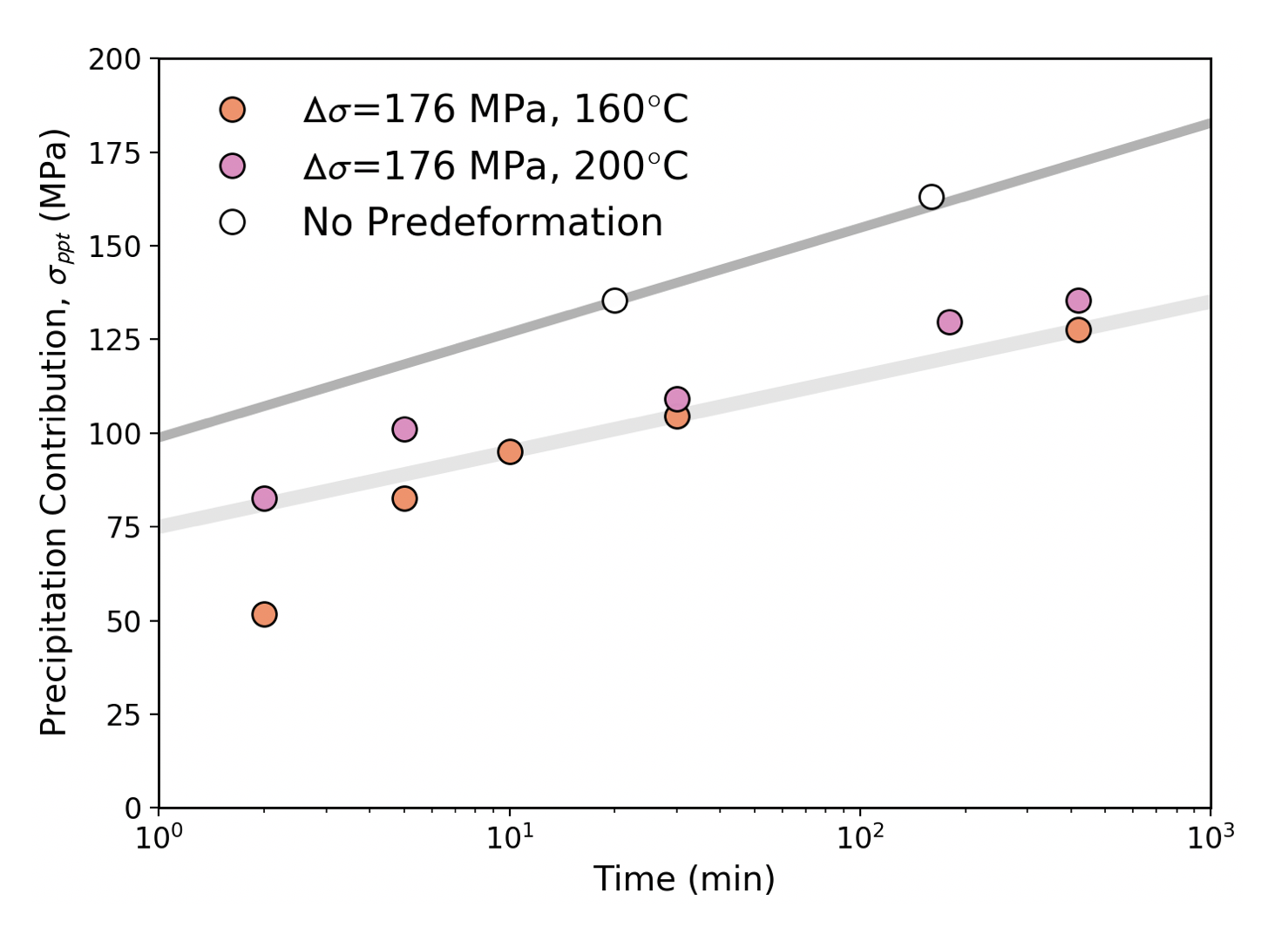}
\caption{The estimated contribution to the flow stress from precipitation hardening, $\sigma_{ppt}$ as a function of aging time and temperature.  Also shown are the results obtained from samples that had been aged without pre-deformation \cite{medrano_transient_2020}. }
\label{fig:precipitationhardening}
\end{figure}

Figure \ref{fig:precipitationhardening} shows the predicted evolution of $\sigma_{ppt}$ with aging time and temperature for these samples.  Also shown here are results from the same alloy aged at 200$^\circ$C from the solution treated state \cite{medrano_cluster_2018}.  Coinciding with the results in Figure \ref{fig:yieldstrength} we see that, except for short aging times at 160$^\circ$C, the aging response follows the same trend regardless of aging temperature.  This also mirrors the aging response that was reported for the as-solution treated samples \cite{medrano_cluster_2018}.  What is notable here, however, is the fact that the precipitation contribution to the flow stress is \emph{lower}, compared at equal aging time, for the samples that had been pre-deformed compared to those that were aged starting from the as-solutionized state. This would suggest that pre-deformation leads to less efficient precipitation hardening.

\begin{figure}[htbp]
\centering
\includegraphics[width=\figwidth\textwidth]{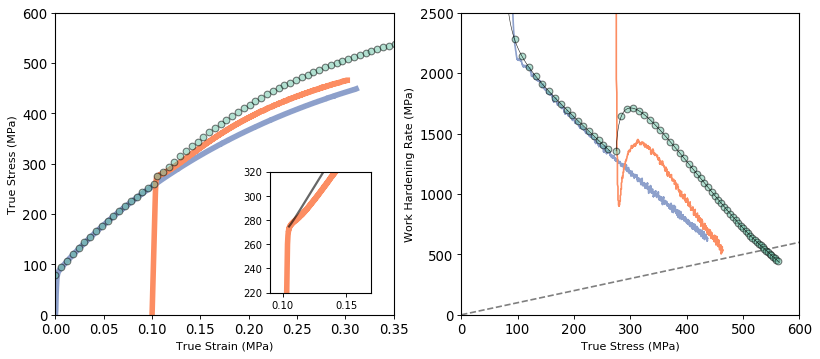}
\caption{The a) stress-strain and b) work hardening response for a sample aged at 200$^\circ$C for 420 minutes.  The model results (symbols) have been obtained using the same values of $\rho_f$ and $\rho_f^*$ as for shorter times and a value of $\sigma_{ppt}$ = 85 MPa, this being in line with the trend shown in Figure \ref{fig:precipitationhardening}.  It is clear, however, that the model cannot reproduce the shape of the stress strain (work hardening rate) in this case.  This can be linked to the fact that the precipitates are likely non-shearable (S-phase) by this stage of aging. }
\label{fig:overprecipitationhardening}
\end{figure}

While the results in Figures \ref{fig:stressstrain_workhardening} and \ref{fig:illustrate_fitting} indicate a very good ability of the model to reproduce the experimental stress-strain response, this was not true for all cases.  In particular the sample aged for the longest time (420 min) at 200$^\circ$C could not be satisfactorily fit by the model.  Figure \ref{fig:precipitationhardening} shows the experimental results for this condition.  It is notable here that the yield stress of the pre-deformed and aged sample is slightly higher than the flow stress of the material at the end of pre-deformation.  Given that recovery has happened, this suggests significant precipitation hardening.  However, if one looks at the flow-stress, work hardening plot (Figure \ref{fig:overprecipitationhardening}b) one sees that the slope of the work hardening rate of the aged sample at high flow stress does not approach parallel with that of the solution treated sample.  Indeed, the two curves appear to cross close to the point of necking.  Given the arguments made above regarding the effect of $\sigma_{ppt}$ this would seem inconsistent.  

This observation is a likely consequence of a majority of precipitates no longer being shearable.  Indeed, in our detailed work on the precipitation hardening in the absence of pre-deformation \cite{medrano_cluster_2018} it was found that a significant portion of the observed precipitates were above the estimated shearable/non-shearable transition after aging for 160 minutes at 200$^\circ$C.  With further over-aging of the sample, the precipitates will influence not only the yield strength but also the work hardening rate of the material, these effects not being properly accounted for in the model used here. 

\subsection{Mechanisms Contributing to The Mechanical Response}

The work detailed above shows that the evolution of the yield strength with aging is dominated by precipitation hardening in the case of pre-deformed samples.  Recovery occurs rapidly at short times but then stagnates once significant solute segregation or precipitation on dislocations occurs.  In this case, the ability of small Cu additions to reduce the strength loss on aging would appear to be a combined effect arising both from precipitation strengthening and the cessation of dislocation loss via slowed recovery.  The last question to examine is why the precipitation hardening appears to be slower in the case of pre-deformed samples.

It is believed that vacancies play a vital role in determining the formation of solute clusters in the early stages of aging in Al-Cu-Mg and Al-Mg-Cu alloys (see e.g. \cite{zurob_model_2009,ringer_precipitation_1996,ringer_origins_1997,Marceau_2006}). In the as-quenched state the microstructure can be described as being composed of a random solid solution with a low density dislocations. Quenched in excess vacancies from solution treatment will allow for accelerated formation of clusters due to accelerated diffusion of solute \cite{aaronson_mechanisms_2010,hutchinson_quantitative_2014,deschamps_situ_2011}, with some vacancies becoming incorporated into the clusters/GPB zones, possibly to thermodynamically stabilize small clusters \cite{kovarik_ab_2012, gault_nexus_2017, zurob_model_2009}. Those excess vacancies not captured within clusters/GPB zones may annihilate at other defects (dislocations and/or grain boundaries) or they may cluster to form dislocation loops \cite{bolling_vacancy_1972}. STEM observation on solution treated and aged samples revealed classical helical dislocations and small dislocations loops in the alloys studied here, similar to observations made in commercial Al-Cu-Mg alloys \cite{marceau_solute_2010,shih_precipitation_1996}. The loss of these excess vacancies would be expected to i) slow the formation/growth of clusters/GPB zones due to reduced diffusivity and ii) lower the rate of cluster/GPB-zone nucleation. Such an explanation would account for the `rapid hardening effect' observed for the present alloy, aged from the undeformed state \cite{medrano_cluster_2018}.  

If the solution treated material is pre-deformed (here at 77~K) then one can have the further generation of deformation induced vacancies \cite{mecking_effect_1980}.  Taking parameters typical of Al-Mg alloys and considering the relatively small levels of pre-deformation used here, we expect only a small change in the concentration of vacancies due to our pre-deformation \cite{medrano_study_2018}.  Upon aging, the presence of dislocations can have a significant effect on reducing the vacancy concentration. Dislocations act as sinks for vacancies and the much higher density of dislocations in the pre-deformed sample compared to the as-solutionized sample leads to a potential competition between vacancy annihilation and vacancy incorporation into clusters/GPB-zones. The recent work of Fischer \emph{et al.} \cite{fischer_modeling_2011} has shown the importance of dislocation density on the rate of vacancy annihilation. We find, using the Fischer model parameterized for Al-Mg alloys \cite{medrano_study_2018}, that samples containing a low dislocation density ($\rho \approx 10^{10} - 10^{11}$ m$^{-2}$, consistent with a solution treated sample) will retain a large fraction of their excess vacancy concentration over aging times of minutes.  In contrast, samples containing $\rho \approx 10^{13}-10^{14}$ m$^{-2}$ (consistent with the dislocation densities in the pre-deformed samples studied here) would lose their excess vacancies within seconds, potentially even during the process of heating to the aging temperature. Thus, it is proposed here that the origins of the reduced precipitation hardening for pre-deformed samples arises from the rapid loss of excess vacancies due to the higher density of sinks. 

\begin{figure}[htbp]
\centering
\includegraphics[width=\figwidth\textwidth]{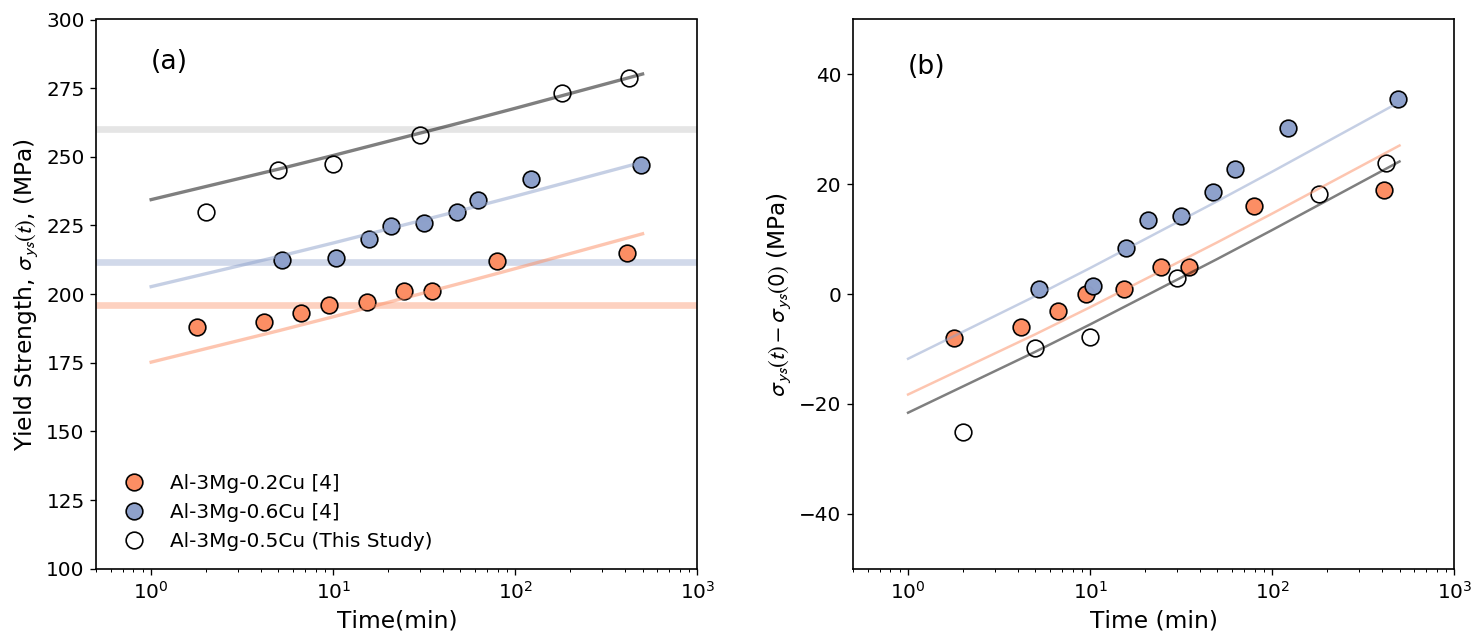}
\caption{a) The evolution of yield strength with aging time collected by Court and Lloyd \cite{court_ageing_2002} for Al-3Mg-0.2Cu and Al-3Mg-0.6Cu (in wt.\%) alloys pre-strained to 10\% strain in tension at room temperature then aged at 180$^\circ$C.  Also plotted are the results obtained here for an Al-3Mg-0.5Cu alloy pre-deformed to $\Delta\sigma = 176$ MPa ($\sim$ 10\% strain) at 77~K). The horizontal lines show the yield strength of the three alloys directly after pre-deformation (without aging)  b) The same data as in (a) but where the yield strength of the pre-deformed material (with no aging) has been subtracted.  The solid lines use the trend line for $\sigma_{ppt}$ shown in Figure~\ref{fig:precipitationhardening}}.  
\label{fig:compare_courtlloyd}
\end{figure}

Finally, reflecting back on the original aim to distinguish between precipitation hardening and recovery induced softening on the yield strength evolution in these alloys, one may ask the question of whether the results presented here are representative of conditions more typical to those found in the automotive paint bake cycle.  Here, for example, deformation was performed at 77~K so as to explicitly avoid the challenges associated with dynamic strain aging at room temperature.  We can, however, compare the results obtained here to those reported by Court and Lloyd \cite{court_ageing_2002} who studied very similar alloys (containing 3wt.\%Mg and 0.2 and 0.6wt.\%Cu) pre-strained in room temperature tensile tests to 10\% plastic strain (very similar to the pre-deformation to $\Delta \sigma =$ 176 MPa used here). Comparing first the yield strength data from their study (Figure~\ref{fig:compare_courtlloyd}a) we see that the yield strength of the sample containing 0.6wt.\%Cu is higher than that of the sample containing 0.2wt.\%Cu. This, however, is complicated by the fact that the yield strength following pre-deformation was slightly higher than the latter.  When these are compared with the 0.5wt.\%Cu alloy studied here, we see that rather than lying between these two, the 0.5wt.\%Cu alloy is situated at much higher stresses.  This, however, arises from the much higher flow stress obtained, despite the similar level of pre-deformation, due to the much lower temperature of deformation (77~K vs. room temperature).

To approximately remove this difference, the data has been re-plotted as the difference between the time evolving yield stress and the yield stress of the as-pre-deformed sample (before aging) in Figure~\ref{fig:compare_courtlloyd}b.  This clearly collapses the data more closely together.  Reflecting on eqn.~\ref{eqn:taylor2}, we see that the data in Figure~\ref{fig:compare_courtlloyd}b is given by,

\begin{eqnarray}
\sigma_{ys}\left(t\right)-\sigma_{ys}\left(0\right) = \sqrt{\left(\sigma_{ss}^2+\sigma_{ppt}^2\right)} + \left(\sigma_{\rho_f}-\sigma_{\rho_{f,0}}\right) - \sigma_{ss}
\end{eqnarray}

In the case of all three samples, $\sigma_{ss}$ is nearly the same since it is dominated by the Mg content of the alloy.  If we take the term relating to recovery, $\left(\sigma_{\rho_f}-\sigma_{\rho_{f,0}}\right)$ to be a constant (cf. Figure~\ref{fig:recovery}) then the only time evolution should arise from the time evolution of $\sigma_{ppt}$.  In the case of the samples from \cite{court_ageing_2002}, we can assume that amount of recovery is very similar thus, the slightly higher stress for the samples containing 0.6wt.\%Cu likely reflects a slightly higher level of precipitation hardening. In the case of the sample containing 0.5wt.\%Cu such an increase is not seen but this is likely due to the fact that one cannot assume the same amount of recovery.  Owing to the higher flow stress (and thus dislocation density) in the pre-deformed state one would expect a faster rate of initial recovery.

Despite the points made above, the three samples in Figure~\ref{fig:compare_courtlloyd} show remarkably similar results despite quite different chemistries, loading conditions and aging temperature.  Indeed, the lines shown in each case use the trend line for $\sigma_{ppt}$ shown in Figure~\ref{fig:precipitationhardening} to predict the evolution of the yield strength. Overall, this points to the fact that variations in chemistry between 0.2-0.6wt.\%Cu and aging temperatures between 160 and 200$^\circ$C will have little effect on the overall rate of strengthening during the paint baking of these alloys and that this evolution is ultimately predictable with limited experimental validation as long as all of the contributions to the strength are properly accounted for.

\section{Conclusion}

It has been shown that the contributions of precipitation hardening and recovery induced softening in pre-deformed Al-Mg-Cu alloys can be separated based on a careful accounting for their unique contributions to the flow stress and work hardening rate. A recently proposed model that accounts for solute-dislocation interactions has been used here to separate these two contributions showing that the evolution of the yield strength with aging is dominated by precipitation hardening. The effect of recovery induced softening saturates within the first minute of aging.  The magnitude of precipitation hardening was found to be lower than that obtained in solution treated (non-pre-deformed samples), this being attributed to the loss of vacancies annihilating at dislocations. Finally, comparing the results obtained here to results reported in the literature, we see that the beneficial effects of small Cu additions are relatively insensitive to the Cu content (between 0.2 and 0.6wt.\%Cu), the deformation temperature and the aging temperature.  This lack of sensitivity means that one can achieve improved properties without very tight composition control on the minor element (Cu) making theses a good starting point for the development of 'recycling friendly' alloys. Studies broadening this work to investigate a wider range of compositions and other important characteristics (e.g. corrosion resistance \cite{engler_influence_2017}) should be undertaken to better understand the limits of composition flexibility.     

\section{Acknowledgements}
C. Bross, U. Tezins and A. Sturm are acknowledged for their support and the running of the APT and FIB facility at MPIE. H. Zhao acknowledges the Chinese Scholarship Council for her PhD scholarship. S. Medrano acknowledges the Mexican National Council of Science and Technology (CONACyT) for his PhD scholarship. C. W. Sinclair and S. Medrano acknowledge the Natural Sciences and Engineering Research Council of Canada for financial support.  Finally Novelis Inc. is thanked for the provision of materials. 

\bibliography{AlMgCu_hardening}

\end{document}